\documentclass[twocolumn,aps,superscriptaddress,showpacs,prl]{revtex4}
\usepackage[latin1]{inputenc} 
\usepackage[francais]{babel}
\usepackage{graphicx}
\usepackage{url}

\begin{document}

\NoAutoSpaceBeforeFDP 

\title{Electronic Instability in a Zero-Gap Semiconductor: \\ the Charge-Density Wave in (TaSe$_4$)$_2$I}

\author{C. Tournier-Colletta}
\email{cedric.tournier@epfl.ch}
\affiliation{Institute of Condensed Matter Physics, Ecole Polytechnique F\'ed\'erale de Lausanne (EPFL), CH-1015 Lausanne, Switzerland}
\author{L. Moreschini}
\affiliation{Advanced Light Source (ALS), Lawrence Berkeley National Laboratory, Berkeley, California 94720, USA}
\author{G. Aut\`es}
\affiliation{Institute of Theoretical Physics, Ecole Polytechnique F\'ed\'erale de Lausanne (EPFL), CH-1015 Lausanne, Switzerland}
\author{S. Moser}
\affiliation{Institute of Condensed Matter Physics, Ecole Polytechnique F\'ed\'erale de Lausanne (EPFL), CH-1015 Lausanne, Switzerland}
\author{A. Crepaldi}
\affiliation{Institute of Condensed Matter Physics, Ecole Polytechnique F\'ed\'erale de Lausanne (EPFL), CH-1015 Lausanne, Switzerland}
\author{H. Berger}
\affiliation{Institute of Condensed Matter Physics, Ecole Polytechnique F\'ed\'erale de Lausanne (EPFL), CH-1015 Lausanne, Switzerland}
\author{A. L. Walter}
\affiliation{Advanced Light Source (ALS), Lawrence Berkeley National Laboratory, Berkeley, California 94720, USA}
\author{K. S. Kim}\affiliation{Advanced Light Source (ALS), Lawrence Berkeley National Laboratory, Berkeley, California 94720, USA}
\author{A. Bostwick}
\affiliation{Advanced Light Source (ALS), Lawrence Berkeley National Laboratory, Berkeley, California 94720, USA}
\author{P. Monceau}
\affiliation{Institut N\'eel, CNRS/Universit\'e Joseph Fourier, 38042 Grenoble, France}
\author{E. Rotenberg}
\affiliation{Advanced Light Source (ALS), Lawrence Berkeley National Laboratory, Berkeley, California 94720, USA}
\author{O. V. Yazyev}
\affiliation{Institute of Theoretical Physics, Ecole Polytechnique F\'ed\'erale de Lausanne (EPFL), CH-1015 Lausanne, Switzerland}
\author{M. Grioni}
\affiliation{Institute of Condensed Matter Physics, Ecole Polytechnique F\'ed\'erale de Lausanne (EPFL), CH-1015 Lausanne, Switzerland}

\begin{abstract}
We report a comprehensive study of the paradigmatic quasi-1D compound (TaSe$_4$)$_2$I performed by means of angle-resolved photoemission spectroscopy (ARPES) and first-principles electronic structure calculations. We find it to be a zero-gap semiconductor in the non-distorted structure, with non-negligible interchain coupling. Theory and experiment support a Peierls-like scenario for the CDW formation below $T_{\rm CDW} = 263$~K, where the incommensurability is a direct consequence of the finite interchain coupling. The formation of small polarons, strongly suggested by the ARPES data, explains the puzzling semiconductor-to-semiconductor transition observed in transport at $T_{\rm CDW}$.
\end{abstract}
\date{\today}

\maketitle

One-dimensional systems exhibit a rich phenomenology due to their reduced phase space. Typically, the metallic state is unstable against competing broken-symmetry phases such as charge- or spin-density waves \cite{gruner,grio09,monceau}. In the dominant Peierls scenario, these transitions are described as electronic instabilities driven by the nesting properties of the Fermi surface. Alternative models have also been proposed \cite{rice,castroneto,ross01}, and recently even the relevance of the Peierls scenario for real materials has been questioned \cite{mazin,webe11}.

The chain compound (TaSe$_4$)$_2$I is a paradigmatic quasi-one-dimensional (1D) material \cite{gres82}. At $T_{\rm CDW} = 263$~K it undergoes a charge density wave (CDW) transition, accompanied by an incommensurate structural distorsion \cite{fuji84,lee85}. As expected, the instability affects the electronic structure, and clear signatures of the transition are found in the electrical resistivity \cite{wang,maki}, the magnetic susceptibility
\cite{johnston} and the optical response \cite{geserich}. The CDW phase is rather unusual. The main part of the lattice modulation is acoustic and transverse \cite{lee85}, with only a much smaller optical component along the chains \cite{vansmaalen} corresponding to Ta-tetramerization modes \cite{favre}. Besides, structural investigations could not establish the existence of a soft phonon mode \cite{lorenzo}. The electrical resistivity is also puzzling. At low temperature, it exhibits an activation gap $\Delta_{\rm LT}\approx 0.3$~eV \cite{maki}, but the temperature dependence is that of a semiconductor also above $T_{\rm CDW}$, albeit with a smaller $\Delta_{\rm HT}\approx 0.2$~eV \cite{bilusic}.  The ratio $\Delta_{\rm LT}/k_BT_{\rm CDW}\approx12$, much larger than the $3.5$ mean-field value, can be rationalized as an extreme manifestation of 1D fluctuations \cite{lee} or, alternatively, of strong electron-phonon coupling \cite{lorenzo}.
 
The electronic structure of (TaSe$_4$)$_2$I has been explored both by angle-integrated and by angle-resolved photoemission spectroscopy (ARPES). Early results reported the absence of a metallic Fermi edge in spectra of the high-temperature phase \cite{sato85,dard91}, suggesting the possible manifestation of the Luttinger liquid behavior \cite{giamarchi}.
It was later recognized that strong electron-phonon coupling leads to the formation of polaronic quasiparticles, and to a dramatic suppression of the coherent spectral weight \cite{perf01}. Further ARPES data revealed signatures of inter-chain coupling \cite{huf99}
and of the incommensurate CDW periodicity \cite{voit00}.

In this Letter, we present an extensive survey of $k$-space by ARPES, and first-principles calculations of the band structure of (TaSe$_4$)$_2$I. The combination of theory and experiment provides an unprecedented view of the three-dimensional electronic structure of this compound. Surprisingly, (TaSe$_4$)$_2$I is found not to be a metal, but a zero-gap semiconductor. The incommensurate periodicity of the CDW appears as a direct consequence of the finite interchain coupling. The ARPES spectral line shapes, typical of small polarons, suggest a natural explanation for the puzzling semiconductor-to-semiconductor transition observed at T$_{\rm CDW}$ in transport data.

\begin{figure*}
  \includegraphics[width=17cm]{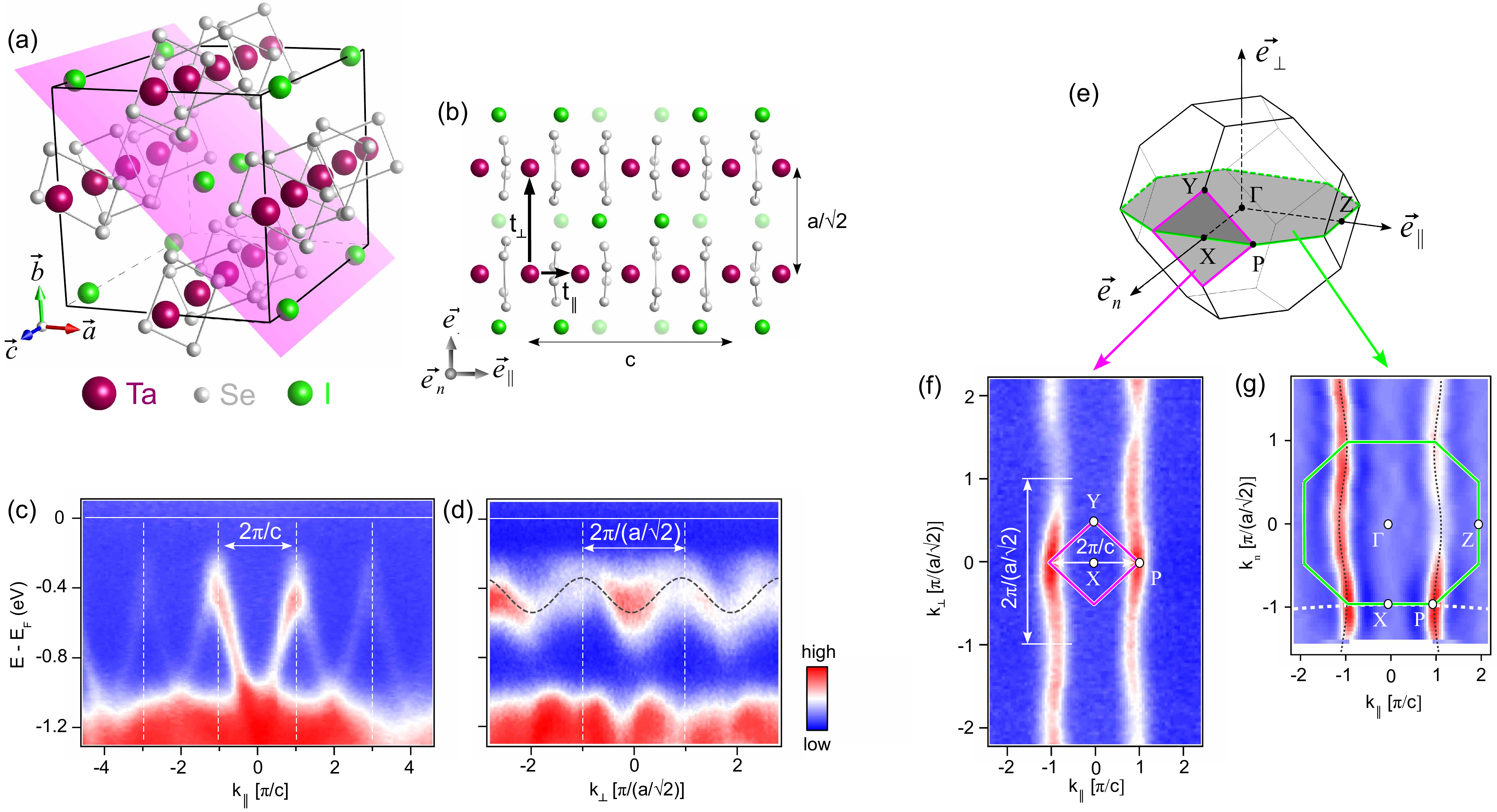}
  \caption{(color online). (a) Crystal structure of non-distorted (TaSe$_4$)$_2$I showing the natural cleavage plane (pink). The conventional tetragonal unit cell is outlined. (b) A projection of the structure onto the cleavage plane. $t_{\parallel}$ and $t_{\perp}$ are the intrachain and interchain nearest-neighbour hopping integrals in the tight-binding model discussed in the text. ARPES intensity maps illustrate the band dispersion (c) parallel (along $XP$) and (d) perpendicular to the chains, for $k_{\parallel} = 0.75~\pi/c$. (e) 3D BZ. (f),(g) Constant energy maps extracted at the top of the Ta $5d_{z^2}$ band ($E_{\rm B} = 0.3$~eV, $T=100$~K) in two perpendicular planes, as depicted in the 3D BZ. Photon energy in (f) is set to $h \nu =88$~eV, that corresponds to the dashed line in (g). Second derivative is used in (g).}
  \label{Fig1}
\end{figure*}

ARPES experiments were performed at beamline 7.0.1 of the Advanced Light Source (Berkeley). The energy and angular resolution of the hemispherical Scienta R4000 analyser were set to $25$~meV and $0.1^\circ$, respectively. The polarization and scattering geometry is described elsewhere \cite{crep12}. We used (TaSe$_4$)$_2$I single crystals grown by the chemical transport method, with typical sizes $5 \times 0.2 \times 0.2$~mm$^3$. They were cleaved {\it in situ} at $100$~K  and a pressure of $10^{-10}$ mbar, and measured at $100$~K, in the CDW phase.
First-principles electronic structure calculations were performed within the density functional theory (DFT) framework using the generalized gradient approximation (GGA). 
Spin-orbit effects were included by means of the fully
relativistic norm-conserving pseudopotentials acting on
valence electron wave functions represented in the two-component
spinor form \cite{cors05}.
We used the Quantum-ESPRESSO software package \cite{gian09}.

Figure~\ref{Fig1}(a) shows the crystal structure of (TaSe$_4$)$_2$I with its conventional tetragonal unit cell ($a=b=9.531$~\AA\ and $c=12.824$~\AA). The Ta atoms form chains, surrounded by Se$_4$ rectangular units and separated by I$^-$ ions. Two adjacent Se$_4$ rectangles are rotated by 45$^\circ$, leading to a periodicity $c = 4 d_{\rm Ta-Ta}$ along the chain. The Bravais lattice actually is not tetragonal, but body-centered tetragonal (bct, space group $I422$ \cite{gres82}). The corresponding rhomboedral unit cell contains two formula units (4 Ta atoms), instead of four (8 Ta atoms) for the conventional cell.  The iodine ions define the natural cleavage plane, perpendicular to $\vec{e}_n = (\vec{a} + \vec{b})/ \sqrt{2}a$. The in-plane unit vectors parallel and perpendicular to the chains are $\vec{e}_{\parallel} = \vec{c}/c$ and $\vec{e}_{\perp}= (\vec{a} - \vec{b})/ \sqrt{2}a$, respectively. The ($\vec{e}_{\parallel}$,$\vec{e}_{\perp}$) and ($\vec{e}_{\parallel}$,$\vec{e}_{n}$) planes are equivalent, reflecting the four-fold symmetry of the space group. Figure~\ref{Fig1}(b) shows the projection of the structure onto the ($\vec{e}_{\parallel}$,$\vec{e}_{\perp}$) cleavage plane. The interchain distance is $a/\sqrt{2} = 6.739$~\AA.
Figure~\ref{Fig1}(e) shows the bct BZ, which exhibits four-fold symmetry, with equivalent $k_{n}$ and $k_{\perp}$ directions. Reciprocal distances are $\Gamma Z = 2 \pi/c = 0.490$~\AA$^{-1}$ (instead of $\pi/c$ for an isolated chain), $XP = \pi/c = 0.245$~\AA$^{-1}$ and $\Gamma X = \pi/(a/\sqrt{2}) = 0.466$~\AA$^{-1}$. The CDW exhibits eight equivalent domains with incommensurate wave vectors ${\bf \delta q}^{\rm CDW}=(\pm\alpha,\pm\alpha,\pm\gamma)$, where $\alpha = 0.045(2\pi/a)$ and $\gamma = 0.085(2\pi/c)$ \cite{lorenzo}.

The in-plane band dispersion measured by ARPES along and perpendicular to the chains is shown in Figs.~\ref{Fig1}(c) and \ref{Fig1}(d). 
Along the chains, the $XP$ direction is probed at this photon energy ($h \nu = 88$~eV; see below). We observe the characteristic ``V''-shaped valence band (VB) \cite{voit00} with a minimum at $k_{\parallel}=0$. The band maxima are at $k_{\parallel}=\pm\pi/c =0.245$~\AA$^{-1}$ and $0.3$~eV binding energy. In a simple picture, it represents the fully occupied, lowest of four subbands split from the nominally quarter-filled band, after gaps are opened at the BZ boundary by the $\times4$ in-chain periodicity \cite{gres84}.
This band is derived from Ta $5 d_{z^2}$ orbitals which overlap strongly along the chain, yielding a small effective mass ($0.3~m_e$, where $m_e$ is the bare electron mass) and an overall band width $W_{\parallel} \approx 4$~eV. Strong polaronic effects drastically reduce the coherent quasiparticle (QP) weight. The largest intensity in the ARPES map does not coincide with the QP energy, but rather with the maximum of the incoherent part of the spectral function. Its energy is lower than the QP energy by the polaron binding energy $E_{\rm p} \approx 0.2$~eV \cite{perf01}. Weaker replicas of the VB are visible at larger wave vectors, in higher-order BZs. 

The VB also disperses in the perpendicular ($k_{\perp}$) direction, with periodicity $2\pi/(a\sqrt{2})$ (Fig.~\ref{Fig1}(d)). The band width is $W_{\perp}\approx 0.2$~eV, corresponding to a smaller but finite inter-chain coupling. This is further illustrated by a constant energy (CE) map, measured at the top of the band (Fig. ~\ref{Fig1}(f)). The open contours are the hallmark of a 1D system, while the periodic ondulations reveal, once again, the interchain coupling \cite{huf99}. 
The overall features of the experimental dispersion can be reproduced by a simple nearest-neighbour tight-binding (TB) model. The hopping parameters can be estimated from the band widths. On a rectangular lattice, $W_{t_{\parallel (\perp)}} = 4 t_{\parallel (\perp)}$. We obtain $t_{\perp} \approx 0.05$ eV and $t_{\parallel} \approx 1$~eV, and $(t_{\parallel} / t_{\perp})^2 \approx 400$ in remarkable agreement with the anisotropy of the electrical conductivity \cite{jero82,forr87}.
\begin{figure}
  \includegraphics[width=8cm]{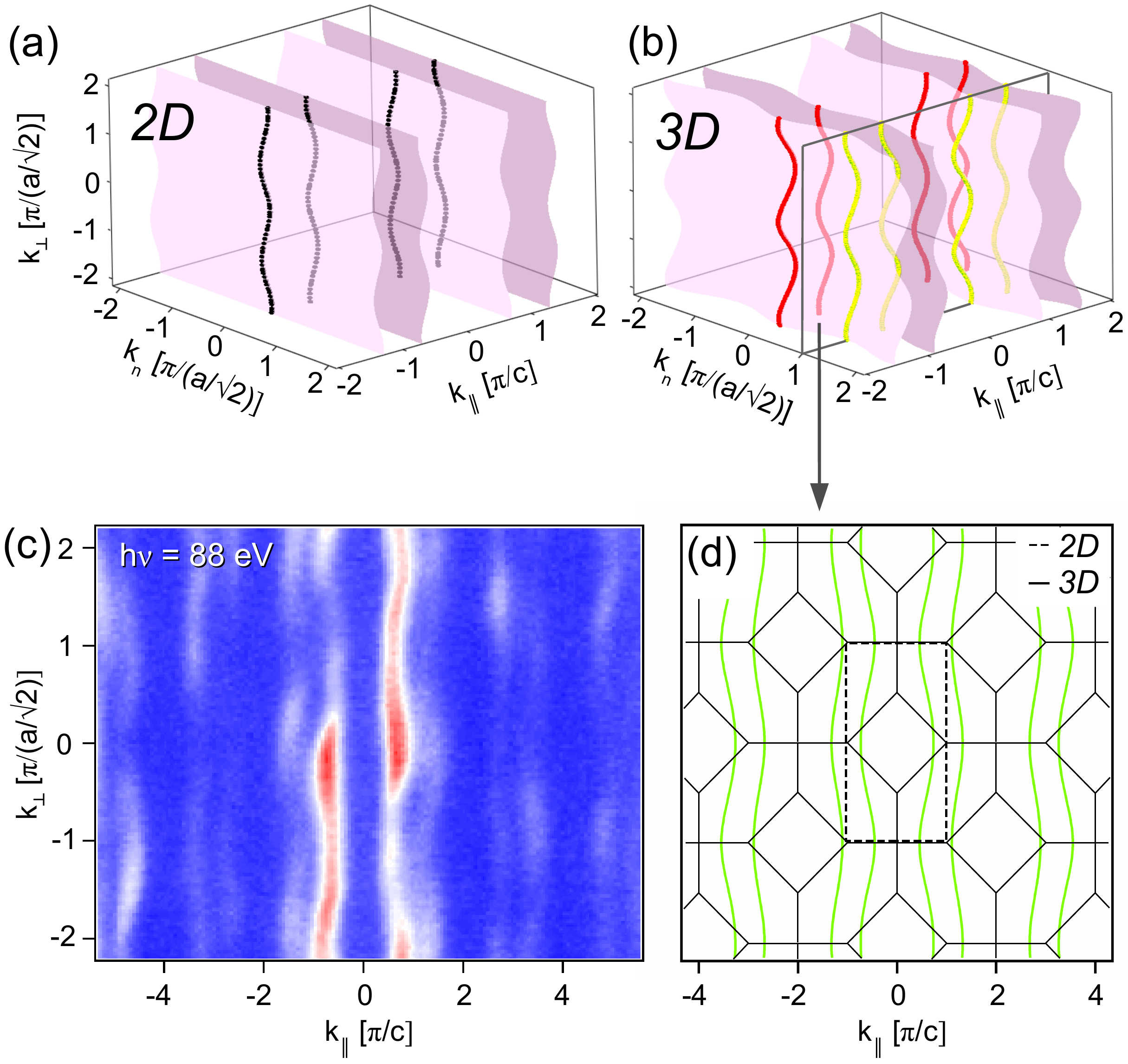}
  \caption{(color online). Schematic constant energy surfaces from (a) a 2D and (b) a 3D tight-binding (TB) model. (c-d) Comparison of experimental CE contours measured at $E_{\rm B} = 0.6$~eV ($h\nu=88$~eV), and theoretical contours derived from a 3D TB model. Projected 3D (solid line) and 2D (dashed) BZs are also indicated }
  \label{Fig2}
\end{figure} 

Figure~\ref{Fig1}(g) reveals a previously unreported dispersion along the $k_n$ direction, perpendicular to the surface. The CE map was extracted from a data set where the photon energy was varied between $80$ and $150$~eV. The inner potential used to determine $k_n$ was $V_0 = 16$~eV. We observe the same wiggling contours as in the $(k_{\parallel}, k_{\perp})$ plane, consistent with the four-fold symmetry of the lattice. As anticipated, the line in $k$ space probed at $h\nu = 88$~eV (white dashed line) almost coincides with the $XP$ direction in the BZ. The separation between the two contours is minimum (``waist'') along this line, and  maximum (``belly''), along the $\Gamma Z$ direction. This difference in the electronic structure between $\Gamma Z$ and $XP$ plays an important role in the CDW formation, as shown below.

In 1D CDW materials, it is often necessary to consider transverse coupling to reproduce the periodicity of the distorted phase. Typically, only one perpendicular direction is relevant. Such an effective 2D scenario, schematically illustrated in Fig.~\ref{Fig2}(a) by CE surfaces obtained from a 2D TB model, is inadequate for (TaSe$_4$)$_2$I.
The two inner sheets correspond to the main band, while the outer two are folded replicas. The $(k_{n},k_{\perp})$ planes at $k_{\parallel} = \pm \pi/c$ define mirror planes between the main band and the replicas. Furthermore, $(k_{\parallel}, k_{\perp})$ CE contours (black lines) do not depend on $k_n$. By contrast, the experimental CE contours of Fig.~\ref{Fig2}(c) exhibit mirror planes at $\pm 2\pi/c = \pm 0.490$~\AA$^{-1}$. An agreement with experiment is found for a 3D TB model [Fig.~\ref{Fig2}(b)]. The constant energy sheets are now warped in both the $\vec{e}_{\perp}$ and $\vec{e}_n$ directions, and the CE contours vary with $k_n$. Indeed from $k_n = 0$ (red) to $\pi/(a/\sqrt{2})$ (green), even if the shape of the contours does not change, the $k_{\parallel}$ separation between opposite branches increases. The calculated contours for $k_n = \pi/(a/\sqrt{2})$ (corresponding to $XP$, Figure~\ref{Fig2}(d)) clearly exhibit the periodicity of the projected 3D BZ (solid line) and not that of the 2D BZ (dashed line). Consideration of the full 3D band structure is therefore essential for a proper description of (TaSe$_4$)$_2$I. This observation provides a rationale for the low anisotropy of the critical fluctuations measured by neutrons above $T_{\rm CDW}$\cite{lorenzo}.
 
\begin{figure}
  \includegraphics[width=8cm]{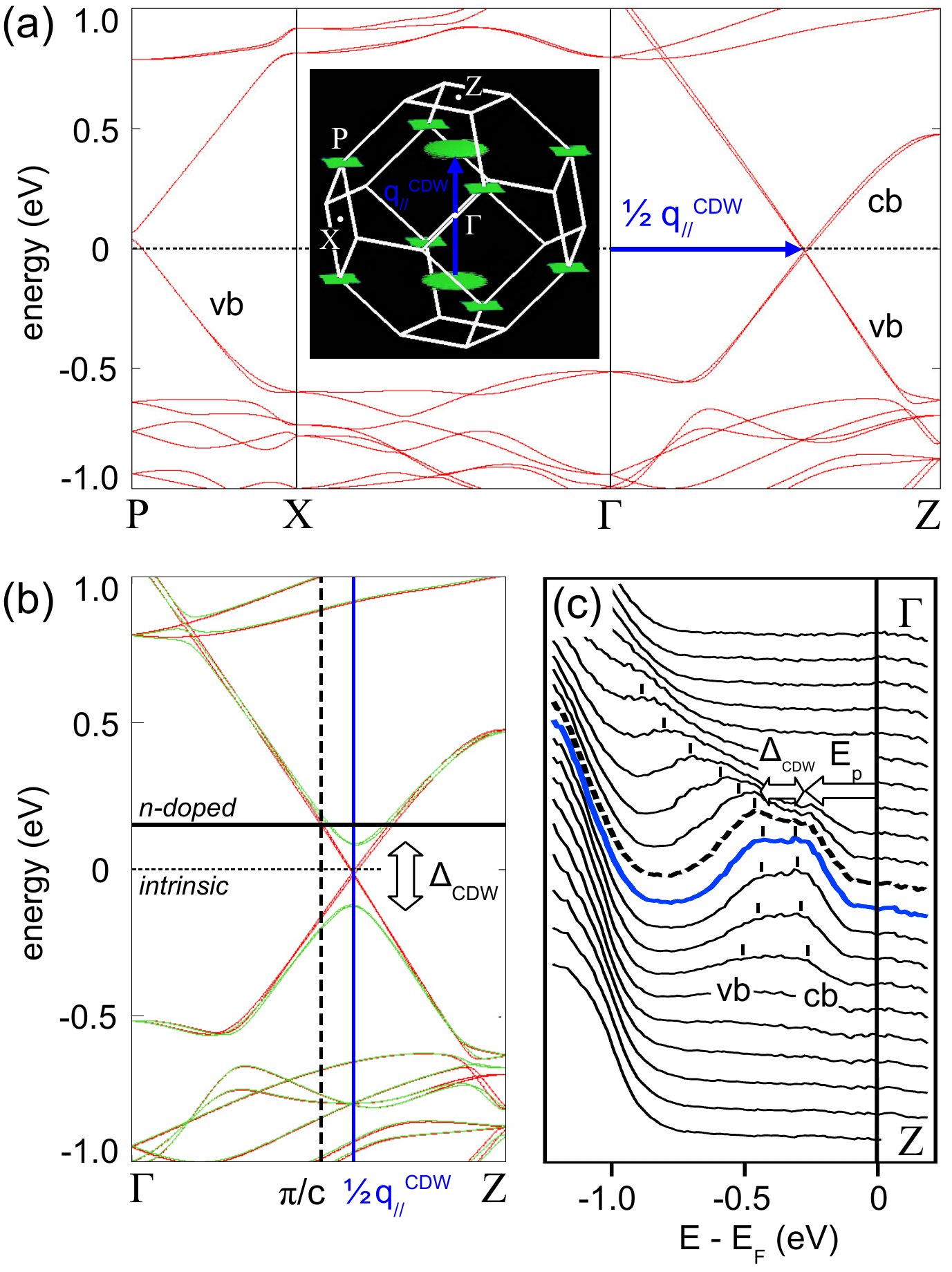}
  \caption{(color online). (a) Electronic band structure of non-distorted (TaSe$_4$)$_2$I calculated from first principles. The inset shows the Fermi surface. The blue arrow depicts the CDW vector. (b) Distorted (green) vs. non-distorted (red) band structures along $\Gamma Z$. The horizontal lines indicate the Fermi level in the intrinsic (dotted) and n-doped (plain) cases. (c) EDCs along the $\Gamma Z$ direction ($h\nu=110$~eV, $T < T_{\rm CDW}$). Natural iodine vacancies yield an effective n-doping that allows to probe CB. Small polaron formation suppresses the QP weight at $E_F$ and transfers the spectral weight at a higher $E_p$ binding energy.}
  \label{Fig3}
\end{figure}

First-principles calculations, performed for the undistorted (high-T) structure, provide important clues on the nature of the instability. The band dispersion is shown in Fig.~\ref{Fig3}(a) for the main high-symmetry directions. Along the $XP$ direction, parallel  to the chains, two $d_{z^2}$-like bands (VB and CB), with extrema at $P$ ($k_{\parallel} = \pi /c$), are separated by a small gap of $\simeq 50$ meV. The Fermi level intersects VB slightly below the top of the band, yielding a narrow hole pocket. By contrast, along the parallel $\Gamma Z$ direction the gap between VB and CB collapses at $E_F$. The top of VB and the bottom of CB merge at $E_F$, yielding a point-like Fermi surface, reminescent of the Fermi surface of graphene. Therefore, the band structure along $\Gamma Z$ is that of a zero-gap semiconductor. The crossing point is not at $\pi/c$, but at the incommensurate $k_F \approx 1.1~ \pi/c$, such that $2k_F\approx q_{\parallel}^{\rm CDW}$. This discrepancy, which reflects the small but noticeable interchain dispersion, has important consequences on the properties of the CDW instability.

Figure~\ref{Fig3}(a) illustrates the calculated Fermi surface (FS) of (TaSe$_4$)$_2$I. The Fermi level was shifted just above the bottom of the conduction band to account for a slight n-doping which is always observed in real samples as a result of iodine vacancies. The FS consists of pairs of very flat `pancakes', perpendicular to the chain direction. The largest pair, along $\Gamma Z$, is nested by the incommensurate CDW wave vector $q_{\parallel}^{\rm CDW}$, consistent with a Peierls scenario. To check that hypothesis, we have performed DFT calculations for the distorted structure, where Ta atoms are mainly displaced along the chains \cite{favre,vansmaalen}. For practical reasons, we considered a commensurate, tetramerized structure  ($q_{\parallel} = 2\pi/c$). The computed band dispersions along $\Gamma Z$ in the high-T (red) and low-T (green) structures are compared in Fig.~\ref{Fig3}(c).  As expected, a gap ($\Delta_{\rm CDW} =0.2$ eV) opens between the valence and conduction bands. Remarkably, the largest energy gain is not at $\pi/c$, but at $k=0.5\times q_{\parallel}^{\rm CDW}>\pi/c$. This observation suggests that the largest gap, and therefore the largest electronic energy gain, is achieved for the actual incommensurate CDW wave vector. Since the elastic energy cost is fairly insensitive to small wave vector changes, these results support the picture of a Fermi surface driven instability. 

The consistency of this scenario is confirmed by the ARPES data of Fig.~\ref{Fig3}(c), which shows energy dispersion curves (EDCs) measured along the $\Gamma Z$ direction in the CDW state. The feature associated with VB indeed reaches a minimum binding energy for $k_{\parallel}^*$ (blue curve) larger than $\pi/c$ (dashed). Moreover, a second peak appears at lower binding energy in a small $k$ range around $k_{\parallel}^*$. It is the signature of the occupied CB states. The experimental CDW gap, defined as the minimum energy separation between the two features, is in good agreement with the theoretical prediction. As for VB, strong polaronic effects are expected to alter the spectral weight distribution of the CB. The measured peak overestimates the actual binding energy by the polaron binding energy $E_p \simeq 0.2$ eV \cite{perf01}. The existence of this additional spectral feature was already suggested by previous data \cite{perf01}, but its true nature was not recognized because a much more limited region of $k$~space was probed in that experiment.

Finally, the proposed scenario naturally explains the puzzling semiconductor-to-semiconductor transition observed in transport at $T_{\rm CDW}$\cite{maki,bilusic}. For $T>T_{\rm CDW}$, the activated behaviour ($\Delta_{\rm HT} \approx 0.2$ eV) reflects the diffusive motion of small polarons, which reduces the conductivity by a factor $e^{-{E_{\rm p}}/{k_{\rm B} T}}$ \cite{mahan}. For $T<T_{\rm CDW}$, the activation energy is increased by the opening of the CDW gap. Transport gives $\Delta_{\rm LT} \approx 0.3$ eV, that compares well to $\Delta_{\rm HT} + \Delta_{\rm CDW} = 0.35-0.4$ eV. In this perspective, estimates of the ratio
$\Delta_{\rm LT}/k_{\rm B}T_{\rm CDW}$ and the importance of 1D fluctations, that act on $\Delta_{\rm CDW}$ only,  should be reconsidered.

In summary, an extensive survey of $k$ space by ARPES, and new first-principles calculations, show that the 1D compound 
(TaSe$_4$)$_2$I is not a metal, but a zero-gap semiconductor with non-negligible transverse interactions. Theory and experiment support a Peierls-like scenario, where the incommensurate CDW vector is a direct consequence of finite interchain coupling. By taking the polaronic nature of QPs, the model also explains puzzling properties of (TaSe$_4$)$_2$I, such as the semiconducting character of the resistivity on both sides of the transition, and the unusually large value of the ratio $\Delta_{\rm LT}/k_{\rm B}T_{\rm CDW}$.


We acknowledge fruitful discussions with J.~E. Lorenzo. This work was supported by the Swiss NSF, namely through grants No. PP00P2\_133552 (G.A. and O.V.Y) and PBELP2-125484 (L.M.). First-principles computations have been performed at the Swiss National Supercomputing Centre (CSCS) under project s336. The Advanced Light Source is supported by the Director, Office of Science, Office of Basic Energy Sciences, of the U.S. Department of Energy under Contract No. DE-AC02-05CH11231. C.~T.-C. and L.~M. equally contributed to this article.

\begin{thebibliography}{30}
\expandafter\ifx\csname natexlab\endcsname\relax\def\natexlab#1{#1}\fi
\expandafter\ifx\csname bibnamefont\endcsname\relax
  \def\bibnamefont#1{#1}\fi
\expandafter\ifx\csname bibfnamefont\endcsname\relax
  \def\bibfnamefont#1{#1}\fi
\expandafter\ifx\csname citenamefont\endcsname\relax
  \def\citenamefont#1{#1}\fi
\expandafter\ifx\csname url\endcsname\relax
  \def\url#1{\texttt{#1}}\fi
\expandafter\ifx\csname urlprefix\endcsname\relax\def\urlprefix{URL }\fi
\providecommand{\bibinfo}[2]{#2}
\providecommand{\eprint}[2][]{\url{#2}}


\bibitem{gruner}
G. Gr\"uner, {\em Density Waves in Solids}~(Addison-Wesley, Reading MA, 1994).

\bibitem{grio09}
M. Grioni, S. Pons, and E. Frantzeskakis, J. Phys.: Condens. Matter {\bf 21}, 023201 (2009).

\bibitem{monceau}
P. Monceau, Adv. Phys. {\bf 61}, 325 (2012).
  
\bibitem{rice}
T. M. Rice and G. K. Scott, Phys. Rev. Lett. {\bf 35}, 120 (1975).

\bibitem{castroneto}
A. H. Castro-Neto,  Phys. Rev. Lett. {\bf 86}, 4382 (2001).

\bibitem{ross01}
K. Rossnagel, O. Seifarth, L. Kipp, M. Skibowski, D. Vo\ss{}, P. Kr\"uger, A. Mazur, and J. Pollmann,  Phys. Rev. B {\bf 64}, 235119 (2001).


\bibitem{mazin}
M. D. Johannes and I. I. Mazin,  Phys. Rev. B {\bf 77}, 165135 (2008).

\bibitem[{\citenamefont{{Weber \textit{et al.}}}(2011)}]{webe11}
\bibinfo{author}{\bibfnamefont{F.}~\bibnamefont{{Weber \textit{et al.}}}},
  \bibinfo{journal}{Phys. Rev. Lett.} \textbf{\bibinfo{volume}{107}},
  \bibinfo{pages}{107403} (\bibinfo{year}{2011}).


\bibitem{gres82}
P. Gressier, L. Guemas, and A. Meerschaut, Acta Cryst. B {\bf 38}, 2877 (1982).

\bibitem{fuji84}
H. Fujishita, M. Sato, and S. Hoshino, Solid State Commun. {\bf 49}, 313 (1984).

\bibitem{lee85}
K. B. Lee, D. Davidov, and A.J. Heeger, Solid State Commun. {\bf 54}, 673 (1985).

\bibitem{wang}
Z. Z. Wang, M. C. Saint-Lager, P. Monceau, M. Renard, M. Gressier, A. Meerschaut, L. Guemas, and J. Rouxel, 
Solid State Commun. {\bf 46}, 325 (1983).

\bibitem{maki}
M. Maki, M. Kaiser, A. Zettl, and G. Gr\"uner, Solid State Commun. {\bf 46}, 497 (1983).

\bibitem{johnston}
D. C. Johnston, M. Maki, and G. Gr\"uner, Solid State Commun. {\bf 53}, 5 (1985).

\bibitem{geserich}
H. P. Geserich, G. Scheiber, M. D\"urrer, F. Levi, and P. Monceau,
Physica B {\bf 143}, 198 (1986).

\bibitem{vansmaalen}
S. van Smaalen, E. J. Lam, and J. L\"udecke, J. Phys.: Condens. Matt. {\bf 13}, 9923 (2001).

\bibitem{favre}
V. Favre-Nicolin, S. Bos, J. E. Lorenzo, J.-L. Hodeau, J.-F. Berar, P. Monceau, R. Currat, F. Levy, and H. Berger,
Phys. Rev. Lett. {\bf 87}, 015502 (2001).
  
\bibitem{lorenzo}
J. E. Lorenzo, R. Currat, P. Monceau, B. Hennion, H. Berger, and F. Levy, J. Phys.: Condens. Matter {\bf 10}, 5039 (1998).

\bibitem{bilusic}
A. Bilusic, I. Tkalcec, H. Berger, L. Forro, and A. Smontara, Fizica A (Zagreb) {\bf 9}, 4 (2000).

\bibitem{lee}
P. A. Lee, T. M. Rice, and P. W. Anderson, Phys. Rev. Lett. {\bf 31}, 462 (1973).

\bibitem{sato85}
E. Sato, K. Ohtake, R. Yamamoto, M. Doyama, T. Mori, K. Soda, S. Suga, and K. Endo, Solid State Commun. {\bf 55}, 1049 (1985).

\bibitem{dard91}
B. Dardel, D. Malterre, M. Grioni, P. Weibel, Y. Baer, and F. Levy, Phys. Rev. Lett. {\bf 67}, 3144 (1991).

\bibitem{giamarchi}
T. Giamarchi, {\em Quantum Physics in One Dimension}~(Clarendon, Oxford, 2003).

\bibitem{perf01}
L. Perfetti, H. Berger, A. Reginelli, L. Degiorgi, H. Hoechst, J. Voit, G. Margaritondo, and M. Grioni, Phys. Rev. Lett. {\bf 87}, 216404 (2001).

\bibitem{huf99}
S. H\"ufner, R. Claessen, F. Reinert, T. Straub, V. N. Strocov, and P. Steiner, J. Electron Spectrosc. Relat. Phenom., {\bf 100}, 191 (1999).

\bibitem{voit00}
J. Voit, L. Perfetti, F. Zwick, H. Berger, G. Margaritondo, G. Gr\"uner, H. H\"ochst, and M. Grioni, Science {\bf 290}, 501 (2000).


\bibitem[{\citenamefont{{Crepaldi \textit{et al.}}}(2012)}]{crep12}
\bibinfo{author}{\bibfnamefont{A.}~\bibnamefont{{Crepaldi \textit{et al.}}}},
  \bibinfo{journal}{Phys. Rev. Lett.} \textbf{\bibinfo{volume}{109}},
  \bibinfo{pages}{096803} (\bibinfo{year}{2012}).

\bibitem[{\citenamefont{{Dal Corso} and {Mosca Conte}}(2005)}]{cors05}
\bibinfo{author}{\bibfnamefont{A.}~\bibnamefont{{Dal Corso}}} \bibnamefont{and}
  \bibinfo{author}{\bibfnamefont{A.}~\bibnamefont{{Mosca Conte}}},
  \bibinfo{journal}{Phys. Rev. B} \textbf{\bibinfo{volume}{71}},
  \bibinfo{pages}{115106} (\bibinfo{year}{2005}).

\bibitem[{\citenamefont{{Giannozzi \textit{et al.}}}(2009)}]{gian09}
\bibinfo{author}{\bibfnamefont{P.}~\bibnamefont{{Giannozzi \textit{et al.}}}},
  \bibinfo{journal}{J. Phys.: Condens. Matter} \textbf{\bibinfo{volume}{21}},
  \bibinfo{pages}{395502} (\bibinfo{year}{2009}).

\bibitem[{\citenamefont{{Gressier \textit{et al.}}}(1984)}]{gres84}
\bibinfo{author}{\bibfnamefont{P.}~\bibnamefont{{Gressier \textit{et al.}}}},
  \bibinfo{journal}{Inorg. Chem.} \textbf{\bibinfo{volume}{23}},
  \bibinfo{pages}{1221} (\bibinfo{year}{1984}).
  
\bibitem{jero82}
D. J\'erome and H. J. Schulz, Adv. Phys. {\bf 31}, 299 (1982).

\bibitem[{\citenamefont{{Forr\'o \textit{et al.}}}(1987)}]{forr87}
\bibinfo{author}{\bibfnamefont{L.}~\bibnamefont{{Forr\'o \textit{et al.}}}},
  \bibinfo{journal}{Solid State Commun.} \textbf{\bibinfo{volume}{62}},
  \bibinfo{pages}{715} (\bibinfo{year}{1987}).
  
\bibitem{mahan}
G. D. Mahan, {\em Many-Particle Physics}~(Plenum Publishers, New York, 2000).

\end{thebibliography}

\end{document}